\documentclass[prb,twocolumn,superscriptaddress,aps]{revtex4-2}
\usepackage[latin1]{inputenc}
\usepackage{graphicx,color}
\usepackage{amsmath}
\usepackage{ulem}

\begin{document}

\title{Vibrational properties of epitaxial graphene buffer layer on silicon carbide}
\author{Guillaume Radtke}
\email[Corresponding author: ] {guillaume.radtke@sorbonne-universite.fr}
\affiliation{Sorbonne Universit\'e, CNRS UMR 7590, MNHN, IMPMC, 4 place Jussieu, 75252 Paris, France}
\author{Michele Lazzeri}
\affiliation{Sorbonne Universit\'e, CNRS UMR 7590, MNHN, IMPMC, 4 place Jussieu, 75252 Paris, France}

\begin{abstract}

The vibrational properties of semiconducting graphene buffer layer epitaxially
grown on hexagonal silicon carbide are determined using
first-principles calculations on a realistic structural model.
Despite the important chemical and structural disorder associated with the partial
covalent bonding with the substrate, the buffer-layer carbon atoms
still display quasi-dispersive phonons mimicking
those of graphene.  The related frequency softening and broadening
provide a natural interpretation of the measured Raman
signal. The vibrations determining thermal conduction are found to delocalize
completely on the SiC substrate, leading to an effective spatial separation 
between material components determining, respectively, electronic and thermal 
transport properties. This situation opens perspectives for thermoelectric applications.

\end{abstract}

\maketitle

The term \textit{buffer layer} (BL) commonly refers to the
first graphitic layer formed after high-temperature silicon sublimation 
from a silicon carbide (SiC) substrate. The BL has been known for decades 
but mainly considered as an unwanted byproduct of the epitaxial growth of 
'proper' graphene, which forms subsequently.
The BL in itself started attracting a specific interest only in recent 
years~\cite{Berger2018,Conrad2017,Nair2017,Zhao2024}.
Indeed, although this atomic layer retains the overall two-dimensional honeycomb 
structure of graphene, the partial covalent interaction with the substrate 
leads to a drastic modification of its electronic structure.
In particular, the observation of a band gap of the order of 0.5~eV
can be a substantial advantage over semi-metallic graphene~\cite{Nair2017}.
Besides, very recently, room temperature hole mobilities exceeding 
5000~cm$^{2}$V$^{-1}$s$^{-1}$ were measured on macroscopic domains of SiC covered 
with this long-range ordered semiconducting carbon layer~\cite{Zhao2024}, 
renewing the interest in the BL as a potential candidate for beyond-silicon 
electronics~\cite{Iacopi2024}. 

There is an overall consensus on the BL atomic structure, 
deduced from low-energy electron diffraction~\cite{Zhao2024,Forbeaux1998,Mallet2007,
Riedl2007,Emtsev2008} and described as a $13\times13$ graphene supercell 
covalently bonded to an underlying Si-terminated SiC 
(0001) surface. In this large cell, known as the $(6\sqrt{3}\times6\sqrt{3})R30^{\circ}$ 
reconstruction and thoroughly 
investigated numerically~\cite{Kim2008,Varchon2008,Deretzis2011,Cavallucci2018},
only a fraction of the BL carbon atoms are bonded to the underlying Si atoms. 
This results in a complex structure where each carbon is surrounded by a different 
environment, providing a locally disordered form of graphene.

Whereas vibrational properties of single and many-layer graphene have
been extensively investigated and are now well understood
~\cite{Ferrari2006,Ferrari2013,Armano2019}, little is known on those of the BL,
in spite of their fundamental character.
Besides, phonons are the main heat carriers 
in semiconductors and play a central role in various applications.
In this context, we stress that the poor thermoelectric properties of graphene are largely 
due to a small Seebeck coefficient, related to its gapless electronic structure, 
but also to a very high thermal conductivity~\cite{Anno2017}. An important 
effort has therefore been devoted to improve its thermoelectric figure of merit, 
through defect engineering or composite design ~\cite{Zong2020,Rathi2024}.
Compared to graphene, the epitaxial BL on SiC displays the advantage of combining 
a sizeable electronic band gap, a high charge carrier mobility and, possibly, a smaller 
thermal conductivity due to the different atomic structure.

The little information available on BL vibrational properties can be attributed, 
experimentally, to the difficulty to grow large and homogeneous areas of high-quality 
material and, theoretically, to the large size of a realistic structure.
Indeed, the reported Raman spectra acquired on BLs display significant 
differences~\cite{Wang2020}, possibly due to samples inhomogeneity, and the few 
theoretical works specifically devoted to the investigation of vibrations used a small, 
unrealistic, $(\sqrt{3}\times\sqrt{3})R30^{\circ}$ reconstruction
to approximate the structure~\cite{Fromm2013,Minamitani2017,Milenov2024}.

In fact, the study of vibrations in the BL/SiC system
requires the implementation of specific techniques beyond the standard ones used 
for \textit{ab initio} lattice dynamics~\cite{Baroni2001,Giannozzi2009}.
As a consequence, a series of relevant questions remain open.
For instance, considering this system as a prototypical form of disordered graphene, 
to which extent do the BL vibrational properties resemble those of pristine graphene ?
What is the origin of the specific Raman signatures visible on the BL spectra ?  
More important, what is the thermal conductivity of the BL
and how does it couple (thermal conductance) with the SiC substrate?
Surprisingly, at present, it is not even clear whether these quantities are meaningful,
given the strong covalent interaction occurring between the BL and the substrate.

\begin{figure*}[t!]
\centering
\includegraphics[scale=0.62]{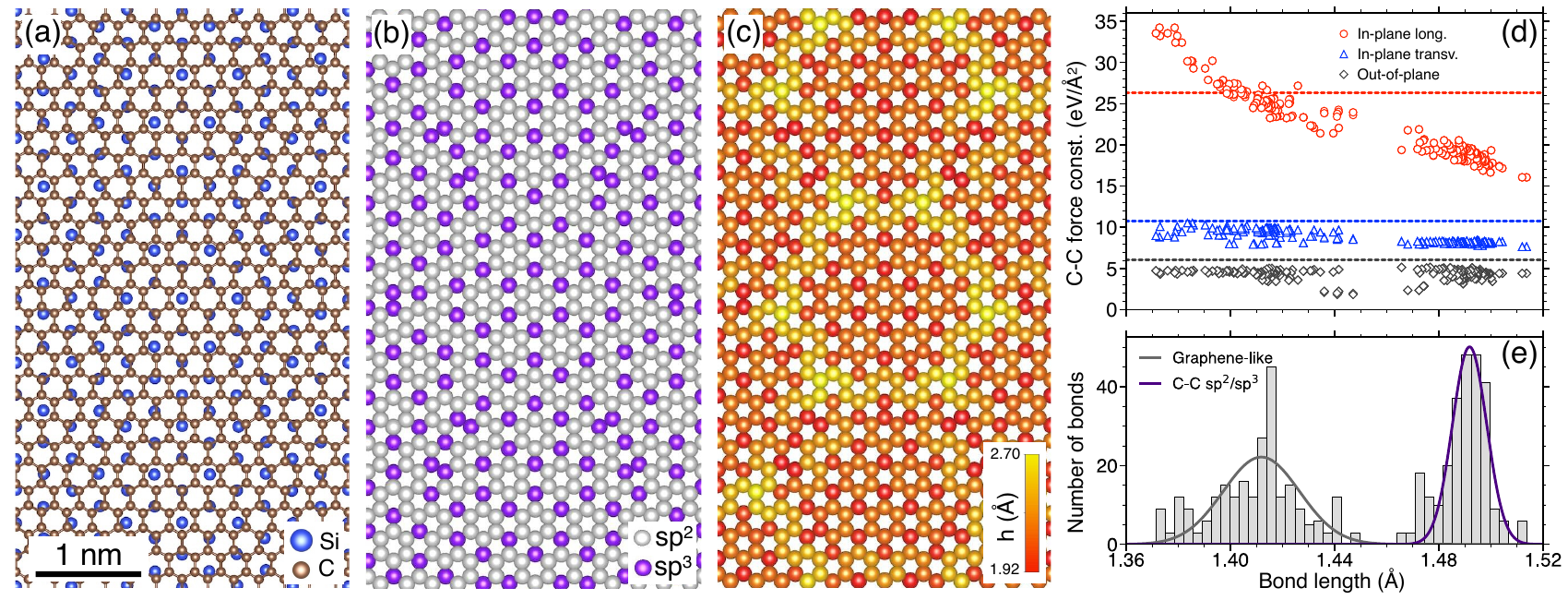}
\caption{(a) Top view of the relaxed atomic 
structure of the $(6\sqrt{3}\times6\sqrt{3})R30^{\circ}$ surface reconstruction. 
Only the carbon atoms of the BL and the topmost Si layer of the substrate is shown.  
(b) Distribution of the $sp^2$ and $sp^3$ hybridized carbon atoms in the BL. 
(c) Height map of the BL carbon atoms.The zero is the average height of the topmost 
Si layer of the substrate. (d) IFCs between nearest neighbors atoms
within the BL calculated for displacements perpendicular (out-of-plane)
or within the atomic plane (in-plane, longitudinal or transverse to the bond direction).
Horizontal lines are the respective quantities calculated for graphene.
The IFCs are reported as a function of the bond length. (e) Histogram distribution of the 
C-C bond length. The two Gaussian distributions correspond to graphene-like
and single C-C $sp^2/sp^3$ group of bonds. }
\label{fig_1}
\end{figure*}

In this Letter, we address these questions by determining fully \textit{ab initio}
the vibrational properties of the BL/SiC system within a realistic structural model.
Calculations were performed using density functional theory~\cite{Giannozzi2009,Baroni2001,
Schlipf2015,Perdew1996}.
Due to the very large system size, a strategy similar to the one developed in 
Refs.~\cite{Hage2020,Bungaro1996,Lazzeri2006} is mandatory: the interatomic 
force constants (IFCs) were first calculated \textit{ab initio} on a structure 
describing the interface and subsequently combined with those obtained from an 
independent calculation of bulk 4H-SiC, to construct the dynamical matrix of a slab 
containing L$_{\rm SiC}$ number of SiC layers. The BL/SiC interface was modeled with 
the $(6\sqrt{3}\times6\sqrt{3})R30^{\circ}$ reconstruction
(\textit{i.e.} 338 carbon atoms, only in the BL).
The vibrational properties of the semi-infinite
substrate were then obtained by adding successive SiC layers until convergence.
While most of the properties are converged for L$_{\rm SiC}=10$ ($\sim 2500$ atoms in total),
calculations up to L$_{\rm SiC}=30$ ($\sim$ 6800 atoms) were necessary.
Extrapolation for ${\rm L_{SiC}}\rightarrow\infty$ is reported for specific properties.
Further details are given as Supplemental Material (SM)~\cite{SM}.

A top view of the BL atomic structure is shown in Figure~\ref{fig_1}(a).
A strong covalent interaction occurs between the BL carbon atoms and the
topmost unsaturated silicon atoms which fall in close coincidence.
Based on both the presence of a short Si-C bond length ($\sim 2$~\AA) and 
a substantial charge density localized between the atoms, about 26~\% of the 
BL carbon atoms are found to adopt an $sp^3$-like hybridization while the remaining 
atoms keep their $sp^2$ configuration, in good agreement with XPS results~\cite{Conrad2017}.
A representation of the spatial distribution of the $sp^2/sp^3$ atoms 
is shown in Figure~\ref{fig_1}(b). The consequences of the electron density 
redistribution within the BL clearly appear in the C-C bond length distribution.
Indeed, Figure~\ref{fig_1}(e) shows two well-separated groups of bonds centered on 1.41~\AA,
close to the equilibrium C-C distance in graphene,
and on 1.49~\AA, representative of typical C-C $sp^2/sp^3$ single bonds~\cite{Allen1987}.
Another consequence is the large deviation of the carbon layer from
the ideal flat structure resulting in a significant rippling of the surface, 
see Figure~\ref{fig_1}(c). The alternation of crests and lower terraces of about 2~nm size
approximately reproduces the $(6\times6)$ periodicity observed in STM 
measurements~\cite{Chen2005,Nair2017} and previous \textit{ab initio} 
calculations \cite{Kim2008,Varchon2008,Deretzis2011,Cavallucci2018}.
While these terraces are tightly bound to the substrate through a high density of 
intruding carbon atoms [in red in Figure~\ref{fig_1}(c)], the crests are essentially 
involving $sp^2$ carbons which do not interact covalently with the substrate.
The resulting average BL/substrate distance is 2.3$\pm$0.2~\AA~in agreement
with transmission electron microscopy~\cite{Palacio2014} and X-ray reflectivity 
measurements~\cite{Hass2008}.

The BL structure is therefore that of an overall two-dimensional hexagonal lattice of
carbon atoms affected by a strong local structural and chemical disorder.
This is reflected in the IFCs, which determine vibrational frequencies. 
For instance, the IFCs between nearest neighbors in the BL reported 
in Figure~\ref{fig_1}(d) are significantly scattered (see also SM~\cite{SM}). 
This is accompanied by a global softening \textit{w.r.t.} to graphene IFCs
of 10~\% for the in-plane longitudinal and up to 30~\% for the out-of-plane components, 
on average. Note that the observed behavior is not trivial.
For example, for the out-of-plane IFCs, it is reasonable to expect that the strong
Si-C covalent interaction would lead to an IFCs softening.
It is however counter-intuitive that this softening
is affecting \textit{all} C-C pairs,
not only those located on the specific sites involving a Si-C bond.

\begin{figure}
\centering
\includegraphics[scale=0.8]{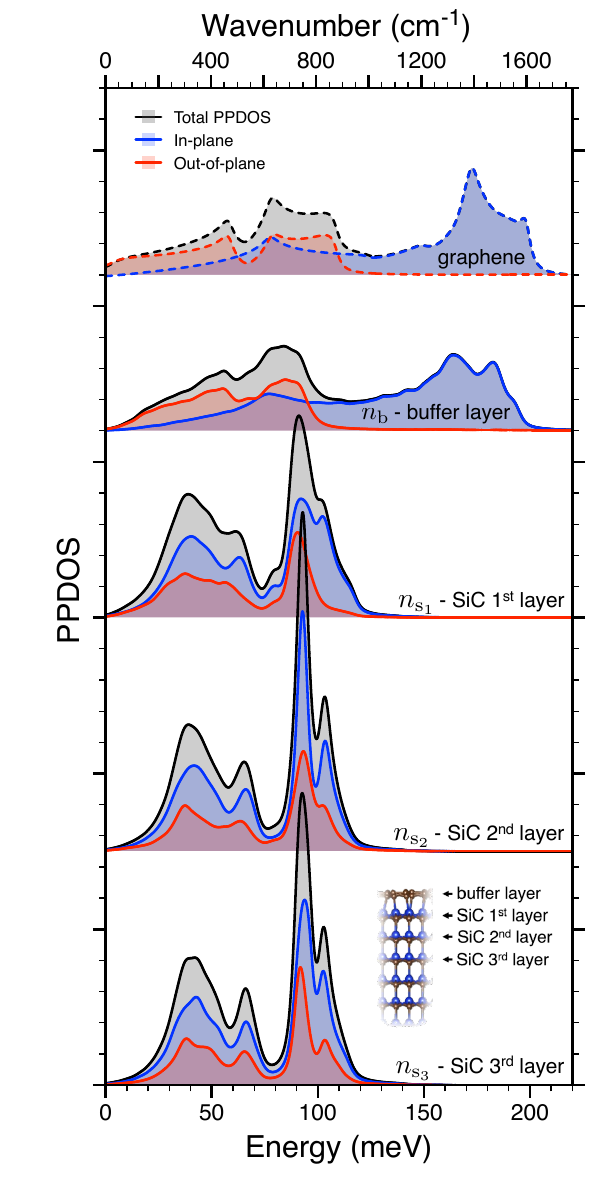}
\caption{Projected phonon densities of states (PPDOS) on the buffer layer ($n_{\rm b}$)
and on the top three SiC layers of the substrate ($n_{\rm s_1}$, $n_{\rm s_2}$, $n_{\rm s_3}$) 
for the BL/SiC interface. As a comparison, the phonon density of states for pristine graphene 
is shown in the topmost panel.}
\label{fig_2}
\end{figure}

We now discuss the projected phonon density of states (PPDOS) associated to different layers
of the BL/SiC interface, since these quantities are nowadays  measurable by electron energy loss
spectroscopy~\cite{Hage2020}. Given a vibrational normal mode $\left.|e_\nu\right>$ with 
frequency $\omega_\nu$, $\left<l,r|e_\nu\right>$ indicates its component on atom $r$ 
of layer $l$. The PPDOS on a layer $l$ is defined as
\begin{equation}
\label{eq1}
n_l(\omega) = \frac{1}{{\rm N_l}{\rm N}_{\boldsymbol{q}}} 
\sum_{\nu,r} |\left<l,r|e_\nu\right>|^2 \delta(\omega-\omega_\nu),
\end{equation}
where the first sum runs over all the modes $\nu$
chosen on a grid of N$_{\boldsymbol{q}}$ wavevectors of the supercell Brillouin zone
and the second one over the ${\rm N_l}$ atoms of the layer (the compound index $r$ 
also includes the three Cartesian components)~\cite{SM}. According to Figure~\ref{fig_2}, 
starting from the second layer, the PPDOS of the SiC is not substantially different
from that of bulk hexagonal SiC, shown in SM~\cite{SM}. The minor differences between the 2$^{nd}$ 
and the 3$^{rd}$ SiC layers are due to their inequivalence in the $4H$ stacking of bulk SiC.
Finally, a pronounced disorder-induced broadening is observed for the topmost SiC layer.
More significant modifications appear when comparing the PPDOS of the BL with the PDOS
of perfect graphene, also reported in Figure~\ref{fig_2}. Although their overall shape 
remains similar, a sizeable broadening and global redshift is observed.
This is particularly visible in the 150-200~meV ($\sim$ 1200-1600~cm$^{-1}$) range of 
optical phonons where a downshift of about 10~meV ($\sim$ 80~cm$^{-1}$) of the 
leading peaks is observed.

\begin{figure*}
\centering
\includegraphics[scale=0.49]{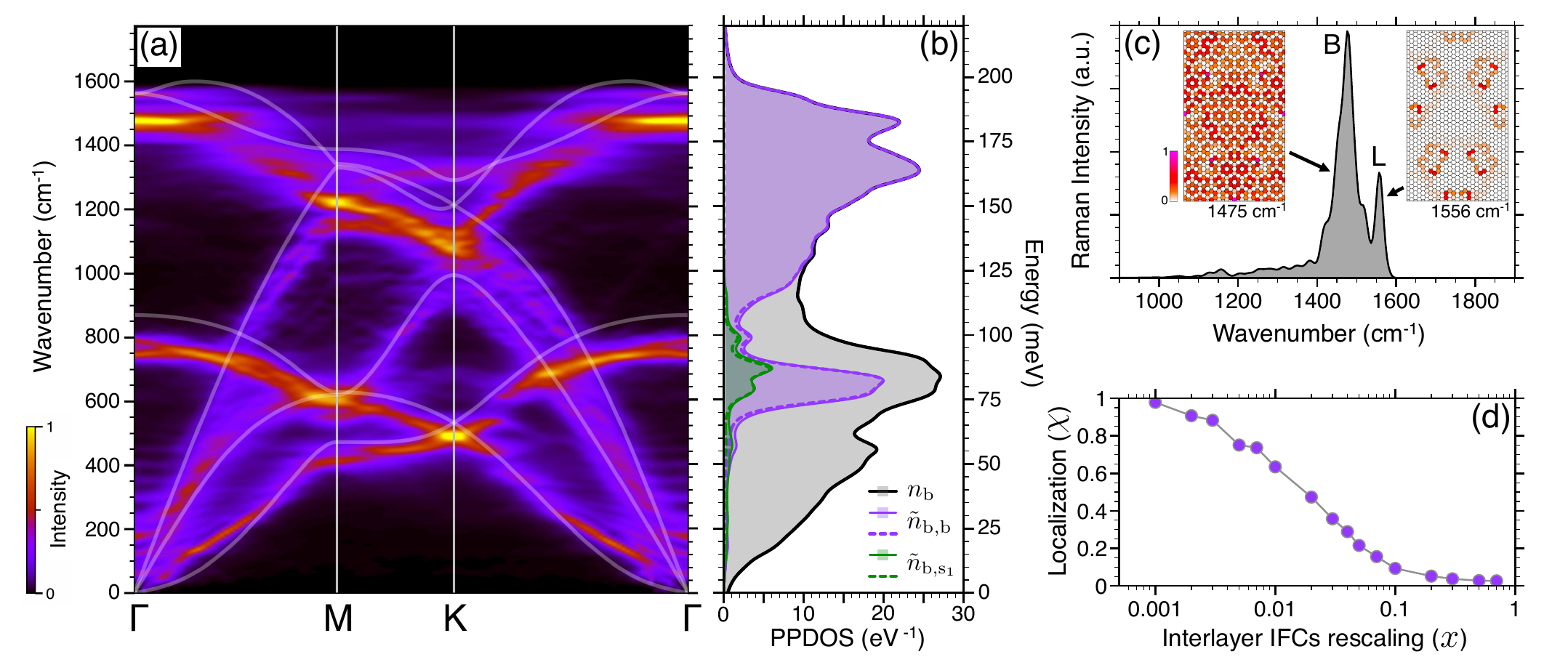}
\caption{(a) Calculated spectral function, $S_{\rm b}(\omega,{\boldsymbol q})$,
for heteroepitaxial BL on SiC along high-symmetry lines of the graphene Brillouin zone.
White lines represent graphene phonon dispersion. (b) Buffer-layer PPDOS, $n_{\rm b}$ from Equation~\ref{eq1},
compared to its decomposition from Equation~\ref{eq2}: $\tilde{n}_{\rm b,b}$ corresponds to $l=l'$
where $l$ is the layer index identifying the BL; for $\tilde{n}_{\rm b,s_1}$ the two indexes $l$ and $l'$ 
identify the BL and the first SiC layer. Continuous lines correspond to a slab with L$_{\rm SiC}$=30. 
Dashed lines are extrapolations to a semi-infinite structure, L$_{\rm SiC}\rightarrow\infty$.
(c) Calculated BL Raman spectrum. Insets indicate the spatial distribution of the atoms 
contributing to the two main peaks (see SM~\cite{SM}). (d) Localization of the acoustic vibrations as 
a function of the inter-layer IFCs ($\chi$ and $x$ are defined in the text).}
\label{fig_3}
\end{figure*}

To gain further insight into these results, we remark that the BL vibrations can be considered as 
those of a disordered 13$\times$13 graphene supercell and, thus, unfolded onto the phonons 
of the 1$\times$1 graphene unit cell. (More precisely, since the vibrational modes are 
defined on the whole BL/SiC slab, we are considering their component
on the atoms of the BL.) Graphene phonons are defined on wavevectors ${\boldsymbol q}$
of the graphene Brillouin zone and the unfolding produces a spectral function
$S_{\rm b}(\omega,{\boldsymbol q})$ (see SM~\cite{SM} for the mathematical definition).
$S_{\rm b}$ reveals to which extent the supercell modes inherit the translational symmetry of the
1$\times$1 crystal: if $S_{\rm b}$ is calculated for a supercell of perfect graphene, it will
provide its phonon dispersion whereas the presence of localized vibrations will result in 
non-dispersive (flat) lines. The comparison of the BL $S_{\rm b}$, presented in Figure~\ref{fig_3}(a) 
as a color-coded map, with the graphene phonon dispersion is interesting in several respects.
Despite the fact that the BL carbon atoms are not arranged in a perfect honeycomb lattice
and that their IFCs present an important degree of local disorder [Figure~\ref{fig_1}(d)],
$S_{\rm b}$ still displays 
characteristics which can be associated to a dispersive two-atom crystal.
Indeed, for a fixed wavevector ${\boldsymbol q}$, the spectral function presents relative maxima at
energies similar to those of graphene and, by varying ${\boldsymbol q}$, these maxima disperse,
in general, in a way resembling that of graphene.
The resulting brilliant stripes in Figure~\ref{fig_3}(a) display however a broadening of 
several meV, which is intrinsic to the BL non-crystalline nature, 
as well as a systematic softening \textit{w.r.t.} graphene
due to the IFCs weakening. 
Note, finally, that the highest optical branch near ${\bf K}$, which for 
graphene/graphite presents a minimum associated to a Kohn anomaly~\cite{Piscanec2004},
becomes flat, an expected consequence of the opening of the 
electronic gap due to the covalent interaction with the substrate~\cite{Nair2017,Cavallucci2018}.

The BL spectral function also determines the one-phonon Raman response.
Indeed, for $\omega>$1000~cm$^{-1}$, $S_{\rm b}(\omega,{\boldsymbol 0})$ projects
the BL vibrations onto the Raman active optical mode of graphene
(${\rm E_{2g}}$, determining the G peak at $\sim$ 1582 cm$^{-1}$~\cite{Ferrari2013}).
The resulting Raman spectrum, shown in Figure~\ref{fig_3}(c),
displays a minor peak at 1556~cm$^{-1}$ (L peak) and a major band at $\sim$1475~cm$^{-1}$ (B peak).
The L peak is due to vibrations localized on very short bonds and depends on the very details 
of the structural model. On the contrary, the B peak is a major feature
of the Raman response, determined by vibrations extending over the entire 13$\times$13 supercell
(more details in SM~\cite{SM}).
As a consequence, a homogeneous $(6\sqrt{3}\times6\sqrt{3})R30^{\circ}$ reconstruction should 
present a well defined B peak whereas the presence of a broad G peak at $\sim$ 1582~cm$^{-1}$ 
should be interpreted as the signature of sample inhomogeneities. 
These findings help rationalize the wide range of spectral features reported for the BL.
Indeed, in a first set of data~\cite{Schumann2014,Strupinski2015,Kruskopf2018}, BL Raman spectra are
only dominated by two broad D ($\sim$ 1350~cm$^{-1}$) and G ($\sim $1580-1600~cm$^{-1}$) peaks
(the D peak is the typical signature of disordered carbon systems~\cite{Ferrari2000}).
In a second set~\cite{Conrad2017,Zhao2024,Bao2016,Kruskopf2016,Wundrack2019,Rejhon2018}, 
the spectra are more structured and display, in addition to D and G, a usually
well-defined B band at $\sim$ 1480-1495~cm$^{-1}$, in close agreement with our calculated frequency.
According to the present results and consistently with a recent experimental analysis~\cite{Wang2020},
these differences should be  attributed to distinct degrees of organization 
in the BL structure, ranging from a disordered one in the first set of spectra to a more 
ordered one in the second set.

At this point, we stress that Figure~\ref{fig_3}(a) results from calculations performed on a 
slab containing several SiC layers, L$_{\rm SiC}$. However, Figure~\ref{fig_3}(a) does not 
tell whether the vibrations contributing to this quasi-phonon dispersion of the BL are actually
localized on the BL itself. In order to investigate this point, we need 
to quantify the localization of the vibrations in the direction perpendicular to the slab surface.
This can be done by generalizing the concept of inverse participation ratio~\cite{Wegner1980}
and defining
\begin{equation}
\label{eq2}
\tilde{n}_{l,l'} (\omega) = \frac{1}{{\rm N}_{\rm l}{\rm N}_{\boldsymbol{q}}} \sum_{\nu,r,r'}
\left| \left< l,r  |e_\nu \right> \right|^2
\left| \left< l',r'|e_\nu \right> \right|^2
\delta(\omega-\omega_\nu)\, .
\end{equation}
When $l'=l$, the comparison of $\tilde{n}_{l,l}$ with $n_l$ can be used to quantify how $n_l$, 
the PPDOS on layer $l$, is due to vibrations that are actually localized 
on $l$: within an energy region where $\tilde{n}_{l,l}\simeq n_l$, $n_l$ is indeed due to vibrations mostly 
localized on layer $l$. On the other hand, since $n_l(\omega)=\sum_{l'} \tilde{n}_{l,l'}(\omega)$,
$\tilde{n}_{l,l'}$ can be used to decompose $n_l$ and to determine how much the modes contributing 
to $n_l$ are localized on a different layer $l'$.

Comparing $n_l$ and $\tilde{n}_{l,l}$ for the buffer layer [$n_{\rm b}$ and
$\tilde{n}_{\rm b,b}$ in Figure~\ref{fig_3}(b)],
we first notice that below 50 meV, for L$_{\rm SiC}\rightarrow \infty$, $\tilde{n}_{\rm b,b}\simeq 0$
meaning that \textit {the acoustic vibrational modes are completely delocalized on the whole
SiC substrate} (see SM~\cite{SM} for more details).
In contrast, the vibrations corresponding to a narrow energy window centered
around 80~meV, strongly relocalize on the BL, as a consequence
of a gap in the phonon spectrum of SiC occurring in that energy range~\cite{SM}.
In the same region, a class of vibrations at $\sim$~87~meV display a component on the
first SiC layer [$\tilde{n}_{\rm b,s_1}$ in Figure~\ref{fig_3}(b)]
and can thus be interpreted as the signature of the BL/substrate covalent bonding.
Note, finally, that all the vibrations
above $\sim$120 meV (in-plane polarized optical modes) are fully localized on the BL
($n_{\rm b}\simeq\tilde{n}_{\rm b,b}$). This obviously results from the absence 
of SiC vibrations in this energy range.

The complete delocalization of BL acoustic phonons over the full SiC substrate, observed
up to $\sim$50~meV, has two important consequences since these vibrations determine 
the thermal and thermal-transport properties up to relatively high temperatures.
The first one is the absence of a relevant interfacial thermal resistance 
between the BL and the SiC substrate (their temperatures cannot differ).
The second one is that 
the substrate will almost entirely determine the thermal conductivity of the system.
In turn, this implies that it is possible, at least in principle, to degrade
thermal transport (for example by defect engineering \textit{within} the substrate)
without affecting the structure of the BL/SiC interface and, thus, the BL itself.
Since the measured high charge-carrier mobility~\cite{Zhao2024}
is likely determined by the electronic states of the semiconducting graphitic layer,
thermal transport can thus be degraded without affecting the electrical conductivity.
Note that, although the thermal conductivity of hexagonal SiC is relatively high,
reaching $\sim$ 370 W/(mK) at room temperature for undoped crystals,
it can be easily degraded~\cite{Slack1964}.

Finally, we remark that the acoustic phonons delocalization is a very robust property 
of the system. Indeed, let us define a localization parameter $\chi$ for 
$\omega<50$~meV vibrations as the normalized integral of $\tilde{n}_{\rm b,b}$ (see also SM~\cite{SM}).
Figure~\ref{fig_3}(d) reports $\chi$ calculated for fictitious systems in which 
the interlayer IFCs between the BL and the substrate are rescaled by a factor $x$.
With respect to the realistic system ($x$=1), where acoustic vibrations are 
delocalized on the substrate ($\chi\sim 0$), a sizable re-localization within the 
buffer ($\chi\sim 1$) is observed only when the interlayer IFCs are reduced to 
few percents of their actual \textit{ab initio} value.

In conclusion, we performed a fully \textit{ab initio} study of the
vibrational properties of the BL/SiC system using a realistic structural 
model including several thousands of atoms.
In spite of being a disordered form of graphene,
the buffer layer still presents vibrations displaying a quasi-dispersive character
resembling that of crystalline graphene. The presence of covalent bonds with 
the substrate results in a overall softening of the vibrations, providing a natural 
explanation of Raman measurements.
Remarkably, as low-energy vibrations are completely delocalized on the SiC substrate,
the thermal conductivity of the system is determined by the substrate itself and, thus, by its quality.
This leads to an effective spatial separation between material components determining,
respectively, electronic and thermal transport properties.
The heteroepitaxial buffer-layer on SiC could then become
a platform for thermoelectric engineering
in the context of the long standing quest for
a system in which the thermal conductivity can be degraded
while preserving the electrical one~\cite{Snyder2008}. \\

We acknowledge support from the French Agence Nationale de la Recherche (ANR) under grants
ANR-23-CE42-0028-01 (project PuMMAVi) and ANR-24-CE50-2146-02 (project BF2D). 
This work was granted access to the HPC resources of IDRIS under the allocations A0100910820R1 
and AD010910820R2 made by Grand Equipement National de Calcul Intensif (GENCI). \\

The data that support the findings of this article are not publicly available. 
The data are available from the authors upon reasonable request.

\end{document}